\documentclass[
twocolumn,
amsmath,
amssymb,
aps,
pra,
]{revtex4} % revtex4-2

\usepackage{graphicx}% Include figure files
\usepackage{dcolumn}% Align table columns on decimal point
\usepackage{bm}% bold math
\usepackage{verbatim}% Ignore the latex command
\usepackage[caption=false]{subfig}% Left-align the caption.
\usepackage{CJK}% Chinese characters can be used
\usepackage{color}

\definecolor{myColor1}{rgb}{1,0,0}
\definecolor{myColor2}{rgb}{0,0,1}
\definecolor{myColor1}{rgb}{0,0,0}
\definecolor{myColor2}{rgb}{0,0,0}
\newcommand*{\newa}[1]{\textcolor{myColor1}{#1}}
\newcommand*{\newb}[1]{\textcolor{myColor2}{#1}}

\begin{document}
\begin{CJK*}{UTF8}{}
\preprint{APS/123-QED}

\title{Critical Exponent of Dynamical Quantum Phase Transition in One-Dimensional Bose-Hubbard Model in the Strong Interacting Limit}% Force line breaks with \\

\author{Jia Li%
\CJKfamily{gbsn}(李佳)}
\affiliation{Department of Physics and Institute of Theoretical Physics, University of Science and Technology Beijing, Beijing 100083, P.~R.~China}%Lines break automatically or can be forced with \\
\author{Yajiang Hao%
\CJKfamily{gbsn}(郝亚江)}
\email{haoyj@ustb.edu.cn}
\affiliation{Department of Physics and Institute of Theoretical Physics, University of Science and Technology Beijing, Beijing 100083, P.~R.~China}
\date{\today}% It is always \today, today, but any date may be explicitly specified

\begin{abstract}
We analytically investigated the dynamical quantum phase transitions in the Bose-Hubbard model using the Loschmidt echo as an observable, revealing that after a quench, the global Loschmidt echo exhibits cusp singularities with a logarithmically divergent rate function near criticality and a critical exponent of zero. Through extensive calculations across various system sizes and initial states, \newa{we have demonstrated that in the strongly interacting regime, the critical singularity of dynamical quantum phase transitions exhibits consistency across different model details and initial product states (charge-density wave states).} Moreover, we find that modifying the harmonic potential well not only preserves the phase transition but also enables precise control over the transition timing.
\end{abstract}

%\keywords{None}% Use showkeys class option if keyword display desired

\maketitle
\end{CJK*}

\section{\label{sec:level1}INTRODUCTION}

Non-equilibrium phase transition is the frontier in quantum physics and statistical physics research, and is also important topics in the fields of condensed matter physics, cold atoms, and quantum control \cite{Cal2011, Cal2016}. In recent years the rapid progress in experiment techniques offers us several powerful platforms for exploring the non-equilibrium dynamics of confined quantum systems such as ultracold atoms \cite{Gro2021}, Rydberg atoms \cite{Bro2020}, superconductivity bits \cite{Kja2020}, and ion traps \cite{Fos2024}. The study of non-equilibrium dynamical quantum phase transitions (DQPTs) is an important part of non-equilibrium phase transition \cite{Hey2013, Hey2018}. DQPTs investigates critical behavior in nonequilibrium quantum systems, whose hallmark feature manifests as non-analyticities in quantum state evolution accompanied by universal critical exponents. These exponents unveil scaling behavior and universality class patterns near quantum critical points. The results align fundamentally with conventional thermodynamic theories, suggesting a high probability of their unification within a single theoretical framework \cite{Hey2014, Kar2013, Kri2014, And2014, Can2014}. The nonanalytic singularities of the Loschmidt rate function at the critical time (cusps in 1D systems) \cite{Hey2015}, or the change of the dynamical topological order parameter \cite{Bud2016}, both indicate the occurrence of the DQPTs. In recent years, groundbreaking research achievements have emerged in this field, driven by advancements in quantum simulation platforms and innovations in theoretical frameworks.

Quantum quenching \cite{Blo2008, Geo2014} has received great attention in the study of non-equilibrium dynamics in isolated quantum systems, especially the Loschmidt echo \cite{Hey2013, And2014, Hic2014, Vaj2014, Sch2015, Vaj2015, Div2016, Bha2017, Kos2018, Lan2018, Lah2019, Liu2019, Cao2020, Zho2021, Zen2023, Cao2023, Kul2023, Sac2024} after the physical conditions suddenly change (the rate of change tends to infinity). Using optical lattices, people have studied the quenching dynamics and phase transitions of quantum many-body systems far from equilibrium \cite{Jep2020, Su2023}. It is worth noting that Loschmidt echo is a global quantity and is not easy to access during the quenching process in the existing experimental schemes, so it is difficult to strictly determine the conditions for the occurrence of DQPTs because the global observables are exponential scale in system size. Therefore, how non-equilibrium quantum phase transitions are reflected in finite-size systems, how to extract and analyze critical times and non-analytical behaviors, that is, how to calibrate kinetic phase transitions in finite systems, are all key issues in DQPTs research.

Recent research has predominantly focused on spin systems, yielding substantial results: DQPTs have been observed both experimentally \cite{Jur2017, Fla2018, Guo2019} and through analytical (numerical) methods, demonstrating universal critical exponents when using either the Loschmidt echo or  translationally invariant Pauli string operator as observables \cite{Ban2021}. Remarkably, consistent phenomena emerge even with finite-length operators when employing the latter approach. These studies have, to a considerable extent, addressed both questions raised earlier. \newa{On the other hand, as another important model in many-body quantum physics, dynamical quantum phase transitions in the Bose-Hubbard model have also been discussed in several theoretical and experimental studies, primarily focusing on quench processes under different model parameters \cite{Stu2022, Lac2019, Abd2019} and improvements in detection methods \cite{Zha2024, kar2025}. Compared with research on integrable models in spin systems, it can be observed that studies of DQPTs in bosonic systems lack precise characterization of the critical singular behavior of observables-primarily the Loschmidt echo. Therefore, calculating the critical exponents at the phase transition is the core objective of this work.}

In this work, we extend the analytical method-originally developed for ground-state dynamical calculations-to product states, employing the Loschmidt echo as the primary observable to investigate DQPTs in the Bose-Hubbard model-this approach enables precise characterization of ultrafast dynamics. \newb{Considering that ultracold atomic gases represent one of the most important platforms for studying quantum systems-particularly given how optical lattice technologies have enabled cold atoms to excel at simulating both the ground state and dynamical properties of quantum many-body systems-we believe it is essential to begin with a theoretical investigation. In this work, we primarily demonstrate the following two points: (i) In the strongly interacting regime, DQPTs in the Bose-Hubbard model exhibit consistent singularities across different initial conditions; (ii) The phase transition time can be controlled by adjusting the external potential.}

\section{Model and method}

The second quantized Hamiltonian for a system of $N$ bosons confined in an optical lattice of length $L$, with only nearest-neighbor tunneling and on-site interactions taken into account, can be expressed as:
\begin{equation*}
\begin{aligned}
    {\hat H} =&  - J\sum\limits_{l = 1}^{L - 1} {\hat b_l^\dag {{\hat b}_{l + 1}}}  + h.c. + \frac{U}{2}\sum\limits_{l = 1}^L {{{\hat n}_l}({{\hat n}_l} - 1)} \\
    & + V_0 \sum\limits_{l = 1}^L {{{\left( {l - \left( {L + 1} \right)/2} \right)}^2}{{\hat n}_l}} ,
\end{aligned}
\end{equation*}
where $\hat b_l^\dag$ and $\hat b_l$ are the creation and annihilation operators for bosons at the lattice site $l$, and ${\hat n_l}$ is the particle number operator. $J$ and $V_0$ denotes the tunenling strength between nearest-neighbor sites and the strength of the harmonic trap, with $U$ represents the on-site interaction strength. Throughout the present paper, we set $J=1$ as the energy unit and the strong interaction limit corresponding to $U \to \infty $ will be investigated. The total number of particles is conserved, satisfying $N = \sum\limits_{l = 1}^L {{{ n}_l}}$. The creation and annihilation operators for bosons, $\hat b_l^\dag$ and $\hat b_l$, obey commutation relations:
\begin{equation*}
\begin{aligned}
    \left[ {{{\hat b}_j},{{\hat b}_l}} \right] \equiv {\hat b_j}{\hat b_l} - {\hat b_l}{\hat b_j} = 0,
\end{aligned}
\end{equation*}
\begin{equation*}
\begin{aligned}
    \left[ {{{\hat b}_j},\hat b_l^\dag } \right] \equiv {\hat b_j}\hat b_l^\dag  - \hat b_l^\dag {\hat b_j} = {\delta _{jl}}.
\end{aligned}
\end{equation*}

The strong interacting limit restricts the additional condition $\hat b_l^2 = \hat b_l^{\dag 2} = 0$. Moreover, we can map the system to spinless fermions using a Jordan-Wigner transformation:
\begin{equation*}
\begin{aligned}
    {\hat b_l} = \exp \left( {{\rm{i}}\pi \sum\limits_{1 \le s < l} {\hat f_s^\dag } {{\hat f}_s}} \right){\hat f_l},
\end{aligned}
\end{equation*}
\begin{equation*}
\begin{aligned}
    \hat b_l^\dag  = \hat f_l^\dag \exp \left( { - {\rm{i}}\pi \sum\limits_{1 \le s < l} {\hat f_s^\dag } {{\hat f}_s}} \right).
\end{aligned}
\end{equation*}

By diagonalizing the single-particle Hamiltonian while neglecting the on-site interaction term, we can obtain the ground-state wavefunction of a spinless free fermion system \cite{Hao2012, Hao2009, Cai2011}, which can be expressed as:
\begin{equation*}
\begin{aligned}
    \left| {\Psi _F^G} \right\rangle  = \prod\limits_{n = 1}^N {\sum\limits_{l = 1}^L {{P_{ln}}} } \hat{f}_l^\dag |0\rangle ,
\end{aligned}
\end{equation*}
where $L$ denotes the number of lattice sites, and $N$ represents the total particle number. The matrix $P$ can be constructed from the lowest $N$ eigenfunctions, where each column corresponds to a single particle eigenfunction. The time evolution of the selected initial state $\left| {\Psi _F^I} \right\rangle $ follows:
\begin{equation*}
\begin{aligned}
    \left| {{\Psi _F}\left( t \right)} \right\rangle  &= {e^{ - {\rm{i}}t{\hat{H}_F}/\hbar }}\left| {\Psi _F^I} \right\rangle  = {e^{ - {\rm{i}}t{\hat{H}_F}/\hbar }}\prod\limits_{n = 1}^N {\sum\limits_{l = 1}^L {{P_{ln}^I}} } \hat{f}_l^\dag \left| 0 \right\rangle \\
    & = \prod\limits_{n = 1}^N {\sum\limits_{l = 1}^L {{P_{ln}}} } \left( t \right)\hat{f}_l^\dag \left| 0 \right\rangle.
\end{aligned}
\end{equation*}
\newa{Where ${\hat H_F}$ is the Hamiltonian of fermions after the Jordan-Wigner transformation, and $P_{ln}^I$ represents the matrix elements of the initial matrix $P$.} The time evolution results can be obtained by directly applying the time evolution operator to the matrix $P$: ${e^{ - it{\hat{H}_F}/\hbar }}{P^I} = U{e^{ - itD/\hbar }}{U^\dag }{P^I}$. The matrices 
$U$ and $D$ are obtained by diagonalizing $\hat{H}_F$: ${U^\dag }{H_F}U = D$.

The Loschmidt echo is defined as ${\cal L}_L(t) = \mid \langle {\psi _0}|\psi (t)\rangle {\mid ^2}$. Consequently, the Loschmidt echo ${{\cal L}_L}(t)$ at any time $t$ is given by \newa{(In all subsequent calculations, we consistently set $\hbar  = 1$, with the time unit fixed as $\hbar /J$ and not explicitly labeled.)}:
\begin{equation}
\begin{aligned}
    {{\cal L}_L}\left( t \right) 
    & = {\left| {\langle 0|{\hat{f}_l}\prod\limits_{n = 1}^N {\sum\limits_{l = 1}^L {{{\left( {P_{ln}^I} \right)}^\dag }} } \prod\limits_{n = 1}^N {\sum\limits_{l = 1}^L {{P_{ln}}} } \left( t \right)\hat{f}_l^\dag \left| 0 \right\rangle } \right|^2} \\
    & = {\left| {\det \left[ {{P^{I\dag }}P\left( t \right)} \right]} \right|^2}.
    \label{eq2}
\end{aligned}
\end{equation}

We explicitly present the construction of the matrix $P^I$ for product states within this strong-interaction computational framework. The matrix $P^I$ is constructed straightforwardly by converting the product state's one-dimensional array (e.g., $\left| {1010 \cdots } \right\rangle  \to [1,0,1,0 \cdots ]$) into a diagonal matrix followed by removing all-zero columns, which can be formally expressed as:
\begin{equation*}
\begin{aligned}
    P_{ln}^I = \delta \left( {l,{n_l}} \right)\delta \left( {n,\sum\limits_{i = 1}^l {{n_i}} } \right).
\end{aligned}
\end{equation*}

The method for calculating the expectation value of particle number products using the $P$-matrix is given by:
\begin{equation*}
\begin{aligned}
   \left\langle {\prod\limits_i^l {{{\hat n}_i}} } \right\rangle  = \det \left[ {{P_{[i:l]}}P_{[i:l]}^\dag } \right],
\end{aligned}
\end{equation*}
where the notation $[i:l]$ represents a programming syntax indicating the extraction of rows $i$ through $l$ from the matrix, where $P_{[i:l]}^\dag$ should first undergo row selection followed by conjugate transposition.

\newa{In the analytical method we use, both the initial state selection (corresponding to experimental preparation) and the evolution must be performed under strong interaction conditions. Therefore, the initial states we can choose are the charge-density wave (CDW) state and the Mott insulating (MI) state \cite{Cha2024}. However, it is evident that the MI state is the ground state of the Hamiltonian discussed in this paper and undergoes no phase transition during the quench process. Thus, in this work we have selected the following two CDW states with product forms: $\left| {{\psi _0}} \right\rangle  = \left| {101010 \cdots } \right\rangle $ and $\left| {11001100 \cdots } \right\rangle $}. These states share identical total particle number and energy (energy degeneracy occurs exclusively at $V_0 =0$), but their distinct initial configurations will significantly influence the subsequent quench dynamics.  \newa{It is worth mentioning that the CDW state is the ground state of the standard extended Bose-Hubbard model \cite{Cha2024} when both on-site and inter-site interactions are large. In our calculations, the on-site interaction remains infinitely large throughout, while the system undergoes quench evolution from an infinitely large inter-site interaction to zero. Such a quenching protocol has been experimentally realized with high precision-using a two-dimensional superlattice and programmable repulsive box potential to prepare a one-dimensional CDW state, followed by instantaneous modulation of the lattice depth to initiate the dynamics \cite{kar2025}.}

\begin{figure}[tb]
    \centering
    \subfloat{\includegraphics[width=8.5cm,height=7.5cm]{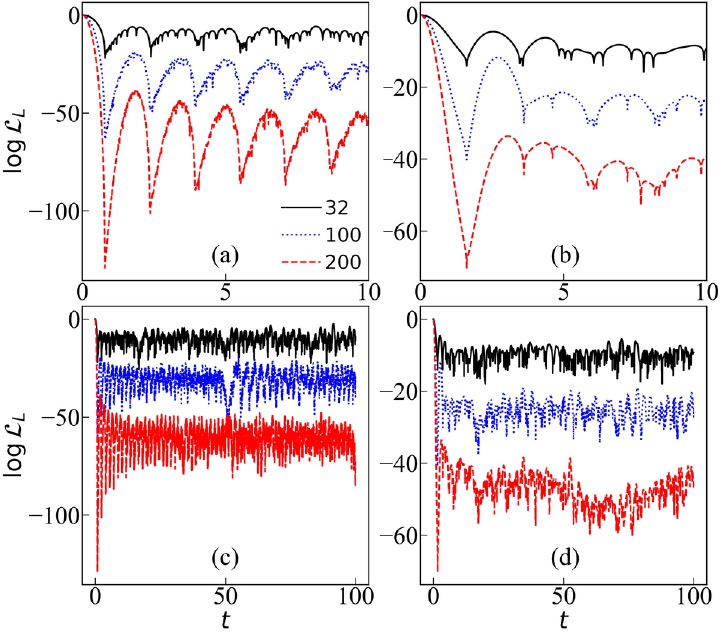}}
    \vspace{0in}
    \caption{Short-time and long-time dynamical properties of the full system ($L = 32,100,200$) are shown for two distinct initial states. All panels sharing a common legend. (a) and (c) $\left| {1010 \cdots } \right\rangle$ with $V_0=0$. (b) and (d) $\left| {1100 \cdots } \right\rangle$ with $V_0=10^{-2}$.}
    \label{fig1}
\end{figure}

Fig. \ref{fig1} demonstrates the short- and long-time evolution of the Loschmidt echo for two distinct initial states, with the latter's timescale being 10 times longer than the former's. Our results demonstrate remarkably similar short-time dynamical evolution between small and large systems for different initial states, with phase transitions occurring at nearly identical times and following consistent critical behavior, where critical point observable in large systems also appear in small systems (Fig. \ref{fig1}a and \ref{fig1}b). We believe these results provide a significant advantage for experimental implementations: the study of DQPTs in smaller systems can yield conclusions similar to those in larger systems, substantially reducing the challenges of system preparation and control. Our long-time dynamics analysis reveals Hilbert space fragmentation, with the ergodic dimension growing exponentially with system size. After crossing the first phase transition point, ${{\cal L}_L}(t)$ rapidly increases and stabilizes within a specific range, establishing a dynamical equilibrium. Our method precisely reconstructs the wavefunction at arbitrary times, providing reliable reference data for experiments and numerical simulations in long-time regimes. \newb{It is worth noting that similar results have emerged experimentally. S. Karch et al. developed a method for detecting DQPTs in subsystems \cite{kar2025}. Comparative analysis shows that the DQPTs behaviors observed in the full system, along with the Hilbert space fragmentation characteristics in long-time dynamics, are largely preserved in the subsystem-a promising finding for developing new experimental approaches.}

The critical behavior of DQPTs near phase transition points is of particular interest in our investigation. The emergent non-analyticity of the rate function $f(t)$ defines DQPTs, expressed in infinite systems as 
\begin{equation*}
\begin{aligned}
   f_L(t) =  - \mathop {\lim }\limits_{L \to \infty } \frac{1}{L}\ln {{\cal L}}_L(t).
\end{aligned}
\end{equation*}
For systems of finite size, the aforementioned definition remains equally applicable. In thermodynamic theory, all thermodynamic functions near critical point can be decomposed into regular parts (finite contributions) and singular parts characterized \newa{by $f\left( t^*  \right) = a{t ^{* \alpha} }\left( {1 + B{t ^{* \beta} } +  \cdots } \right)$. , where $t^* \equiv ({T - {T_c}})/{T_c}$ is the reduced temperature. As a general theory of DQPTs, we treat time as analogous to temperature and define the reduced time $\tau  = \left| {(t - {t_0})/{t_0}} \right|$ in the same manner, where $t_0$ represents the phase transition time. According to thermodynamic theory, the critical exponent $\alpha$ can be defined as}
\begin{equation}
\begin{aligned}
   \newa{\alpha  = \mathop {\lim }\limits_{\tau  \to 0} \frac{{\ln f_L\left( \tau  \right)}}{{\ln \tau }}},
   \label{eq3}
\end{aligned}
\end{equation}
where $\alpha >0$ (e.g., $\alpha=1$ in spin models \cite{Ban2021, Hey2015, Hal2021}) indicates $f_L(\tau)$ vanishing at criticality, whereas we reveal $\alpha=0$ with logarithmic divergence in the Bose-Hubbard model. Through extensive calculations across different initial states \newb{(including but not limited to $\left| {111000 \cdots} \right\rangle $, $\left| {11110000 \cdots} \right\rangle $ and $\left| {1 \cdots 10 \cdots 0} \right\rangle $)}, we conclude that this represents a universal critical exponent in the Bose-Hubbard model. In this work, we focus exclusively on demonstrating the two aforementioned initial states.

\begin{figure}[htbp]
    \centering
    \subfloat{\includegraphics[width=8.7cm,height=9.82cm]{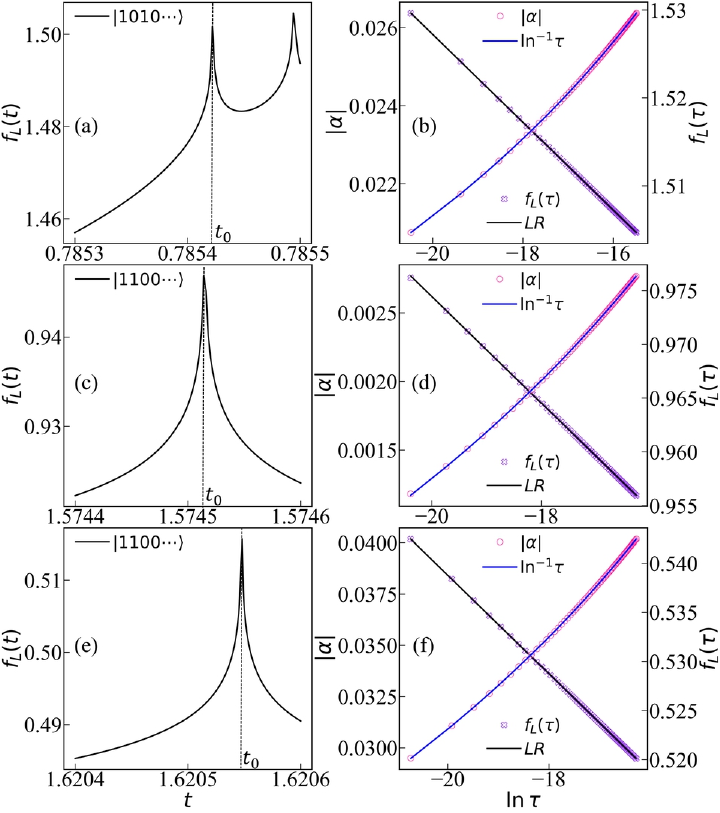}}
    \vspace{0in}
    \caption{The singular behavior of the rate function near the phase transition point is systematically investigated under different conditions: (i) two initial states (panels (a) and (b): $\left| {1010 \cdots } \right\rangle $; other panels: $\left| {1100 \cdots } \right\rangle $); (ii) two potential strengths (panels (e) and (f): ${V_0} = {10^{ - 2}}$; others: 0); with fixed system size $L = 400$. (a), (c) and (e) with timescale $2 \times {10^{ - 4}}$, ${f_L}\left( t \right)$ exhibits a sharply divergent peak at ${t_0}$ (characterized by derivative jumps from $+\infty$ to $-\infty$, distinct from finite jumps in spin systems), whose magnitude increases with temporal resolution-a hallmark of divergence. (b), (d) and (f) with timescale $2 \times {10^{ - 7}}$. All data points represent actual computational results, with the black line showing the linear regression of ${f_L}\left( \tau  \right)$ vs $\ln \tau $, and the blue line indicating the linear regression of $\left| \alpha  \right|$ vs $1/\ln \tau $-both conclusively demonstrating a zero critical exponent. Identical critical exponents are obtained on both sides of the transition.}
    \label{fig4}
\end{figure}

Fig. \ref{fig4} displays the singular behavior of the rate function near the phase transition point, presenting both: temporal evolution $f_L(t)$ versus $t$, and $f_L(\tau)$ versus $\ln \tau$, for two distinct initial states. Near $t_0$, the rate function $f_L(t)$ exhibits anomalously sharp features with consistent behavior across initial states and potential fields (Figs. \ref{fig4}a, \ref{fig4}c and \ref{fig4}e), while near the critical point ${f_L}\left( \tau  \right) \propto \ln \tau$ is evidenced by linear regression fits (correlation coefficient ${R^2} > {\rm{0}}{\rm{.9999}}$, probability value $P < {10^{ - 147}}$) in Figs. \ref{fig4}b, \ref{fig4}d \ref{fig4}f. Eq. \ref{eq3} yields a critical exponent $\alpha$ decaying as $1/\ln \tau $, demonstrating logarithmic divergence of ${f_L}\left( \tau  \right)$ and ultimately $\alpha  = 0$ as $\ln \tau  \to  - \infty $ ($\tau  \to 0$). It should be noted that while we selected a relatively small timescale to best approximate the critical point, the critical singularity of the rate function can in fact be observed at a timescale of ${10^{ - 2}}$.

\newb{In contrast to the two-dimensional Ising model, where the divergence of specific heat with reduced temperature serves as direct evidence for the divergence of the correlation length and the emergence of fluctuations at macroscopic scales, we speculate that at the critical point, within the quantum states orthogonal to the initial state, there may exist divergences or fluctuations of other observables.}

\begin{figure}[htbp]
    \centering
    \subfloat{\includegraphics[width=8.5cm,height=3.98cm]{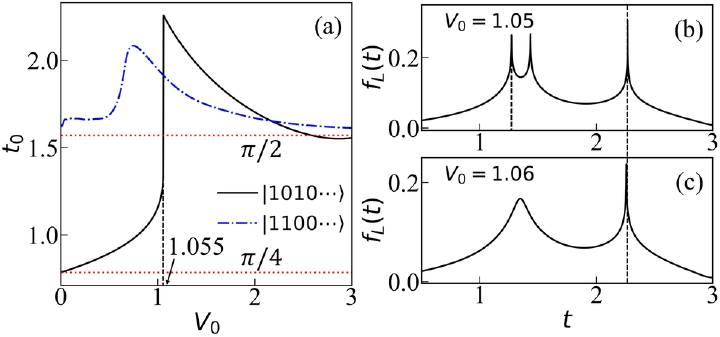}}
    \vspace{0in}
    \caption{(a) The timing of the first phase transition versus potential well strength $V_0$ (occurring for all ${V_0} \ge 0$). (b) and (c) Rate functions for the $\left| {1100 \cdots } \right\rangle $ state at ${V_0} = 1.05$ and 1.06, respectively. The lattice length $L$ was fixed at 32 with a potential resolution of $V_0$ set to 1/300.}
    \label{fig3}
\end{figure}

The timing of the first phase transition presents an intriguing question, as Fig. \ref{fig3}a illustrates its dependence on potential strength. For the $\left| {1010 \cdots } \right\rangle $ state, ${t_0} \approx \pi /4$, while for $\left| {1100 \cdots } \right\rangle $, ${t_0} \approx \pi /2$, with $t_0$ exhibiting non-monotonic dependence on ${V_0}$ initially increasing, then decreasing, and finally asymptotically approaching $\pi /2$ at strong potentials. The behavior at ${V_0} \to  + \infty $ can be qualitatively explained as follows: to conserve energy, particle hopping between potential wells becomes restricted, while the central region with one filled and one empty site (equal potential energy) reduces the system dynamics to an effective two-site problem, yielding the rate function ${f_2}\left( t \right) =  - \ln \cos \left( t \right)$ with ${t_0} = \pi /2$ and $\alpha =0$.

The $t_0$ evolution for the $\left| {1100\cdots} \right\rangle $ state shows monotonic dependence on $V_0$, while the $\left| {1010\cdots} \right\rangle $ state exhibits a sharp discontinuity at $V_0 = 1.055$ (Fig. \ref{fig3}a). Fig. \ref{fig3}b and \ref{fig3}c reveal that when $V_0$ changes from 1.05 to 1.06, the original first and second phase transition points vanish while the third remains stable. These findings suggest potential experimental approaches for controlling DQPTs through external potentials, though the underlying mechanisms require further investigation. \newb{Given that the parameters we used when observing the jump-$L = 32$ and ${V_0} = 1.055$-are relatively small, and the original interval between the two phase transition peaks exceeds 0.1$\hbar /J$, we expect this phenomenon is experimentally observable. By maintaining a strong on-site interaction and scanning the range ${V_0} = 1$ to 1.1, the number of peaks within 2$\hbar /J$ seconds can be counted to determine the jump in phase transition time, without needing to track the detailed behavior of the rate function near the peaks-this therefore substantially reduces the difficulty of observation.}

\begin{figure}[htbp]
    \centering
    \subfloat{\includegraphics[width=8.5cm,height=4.04cm]{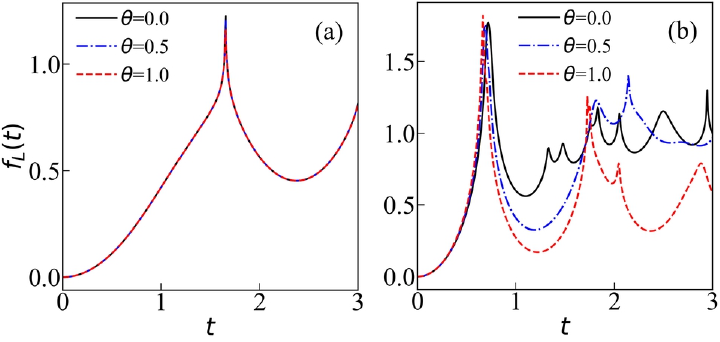}}
    \vspace{0in}
    \caption{Evolution of the $\left| {1100 \cdots } \right\rangle $ state's rate function under different statistical parameters: (a) $U/J=1000$, (b) $U/J=0.1$. Key two-site TDVP algorithm parameters: lattice length $L=20$, maximum occupancy per site $n_{max}=3$, maximum number of singular values $\chi=512$, time step $dt=0.005$.}
    \label{fig2}
\end{figure}

This study further reveals a significant conclusion: \newb{In the anyon Hubbard model \cite{Wil1982, Dha2025, Hao2009}, when studying DQPTs in the strongly interacting regime, the influence of the statistical parameter $\theta$ on the rate function becomes trivial}, as the transformation of bra and ket wavefunctions in Eq. \ref{eq2} mutually cancels out, completely eliminating all phase information from ${{\cal L}_L}\left( t \right)$. Here we employ the time-dependent variational principle (TDVP) algorithm to validate this conclusion through anyon-Hubbard model (The hopping term transforms from $\hat b_l^\dag {\hat b_{l + 1}}$ to $\hat b_l^\dag {\hat b_{l + 1}}{e^{{\rm{i}}\theta \pi {{\hat n}_l}}}$, introducing statistical parameter $\theta$ and occupation number ${{{\hat n}_l}}$.) simulations. Fig. \ref{fig2} compares the rate function evolution for the $\left| {1100 \cdots } \right\rangle $ state in strong (Fig. \ref{fig1}a, $U/J=1000$) and weak (Fig. \ref{fig1}b, $U/J=0.1$) interaction regimes, revealing complete overlap across statistical parameters in the strong-coupling limit versus distinct difference in weak interactions. \newb{Meanwhile, as shown in Fig. \ref{fig2}b, even when the product state evolves in the weakly interacting regime, similar divergence points still emerge in the rate function. However, due to the precision limitations of our numerical methods, this issue will not be further explored in the this work.}

\section{Conclusion}

In summary, we have derived an analytical representation of product states using the $P$ matrix and computed the Loschmidt echo in strongly interacting systems, demonstrating that DQPTs in the Bose-Hubbard model manifest as singularities in the time evolution of the Loschmidt echo. Furthermore, by tracking the behavior of various \newa{initial product states ($\left| {101010 \cdots } \right\rangle $, $\left| {11001100 \cdots } \right\rangle $, $\left| {111000 \cdots} \right\rangle $, $\left| {11110000 \cdots} \right\rangle $ and $\left| {1 \cdots 10 \cdots 0} \right\rangle $ et al.)} near the critical point, we identified a universal critical exponent $\alpha=0$, with the rate function exhibiting logarithmic divergence. This marks the first analytical calculation of a DQPTs critical exponent in the Bose-Hubbard model. Nevertheless, several key questions remain beyond this study's scope: (i) The optimal selection of local operators for DQPT detection (notably challenging due to the peculiarity of critical exponents); (ii) The influence of alternative external potentials or interaction strengths beyond harmonic confinement. Our findings nevertheless establish a robust theoretical foundation, paving the way for future numerical and experimental investigations in this field. \\ \\

\noindent{\bf Data availability} \\
The authors confirm that the data supporting the findings of this study are available within the paper [and/or its supplementary materials].

\bibliography{apssamp}% Produces the bibliography via BibTeX.

\end{document}